\documentclass{IEEEtran}  

\IEEEoverridecommandlockouts                              

\makeatletter
\renewcommand{\@IEEEsectpunct}{.\ \,}
\makeatother

\usepackage{graphics} 
\usepackage{epsfig} 
\usepackage{mathptmx} 
\usepackage{times} 
\usepackage{amsmath} 
\usepackage{amssymb}  
\usepackage[]{units}
\usepackage{gensymb}
\usepackage{placeins}
\usepackage{dblfloatfix}  
\usepackage{arydshln}

\title{\LARGE \bf
In-Ear SpO\textsubscript{2} for Classification of Cognitive Workload}

\author{Harry J. Davies, Ian Williams, Ghena Hammour, Metin Yarici, Barry M. Seemungal and Danilo P. Mandic
}

\begin{document}

\maketitle
\thispagestyle{empty}
\pagestyle{empty}

\begin{abstract}
Classification of cognitive workload promises immense benefit in diverse areas ranging from driver safety to augmenting human capability through closed loop brain computer interface. The brain is the most metabolically active organ in the body and increases its metabolic activity and thus oxygen consumption with increasing cognitive demand. In this study, we explore the feasibility of in-ear SpO\textsubscript{2} cognitive workload tracking. To this end, we preform cognitive workload assessment in 8 subjects, based on an N-back task, whereby the subjects are asked to count and remember the number of odd numbers displayed on a screen in 5 second windows. The 2 and 3-back tasks lead to either the lowest median absolute SpO\textsubscript{2} or largest median decrease in SpO\textsubscript{2} in all of the subjects, indicating a robust and measurable decrease in blood oxygen in response to increased cognitive workload. Using features derived from in-ear pulse oximetry, including SpO\textsubscript{2}, pulse rate and respiration rate, we were able to classify the 4 N-back task categories, over 5 second epochs, with a mean accuracy of 94.2\%. Moreover, out of 21 total features, the 9 most important features for classification accuracy were all SpO\textsubscript{2} related features. The findings suggest that in-ear SpO\textsubscript{2} measurements provide valuable information for classification of cognitive workload over short time windows, which together with the small form factor promises a new avenue for real time cognitive workload tracking.

\end{abstract}


\section{Introduction}

\IEEEPARstart{C}{ognitive} workload is defined as a the level of mental effort undertaken by an individual in response to a task. The mental effort is usually related to working memory, and thus corresponds to the utilisation of brain resources \cite{Ayaz2010}. Cognitive workload affects almost every task-related aspect of our daily lives, from general learning to driving to internet browsing. The ability to accurately measure cognitive load would lead to manifold benefits, as too little cognitive workload leaves us vulnerable to distraction, whereas too much cognitive workload makes us prone to making mistakes. Depending on the task a person is engaged in, these mistakes can be more benign, such as less efficient studying, through to life threatening ones as is the case with driving and the possibility of fatal accidents. The ability to accurately measure and predict cognitive workload would therefore make possible personalised task adaptation together with the associated benefits on an individual and a societal level, from increasing productivity to decreasing the likelihood of mistakes.

\subsection{Cognitive workload tracking}

It is natural to attempt to track cognitive workload based on scalp electroencephalography (EEG), with examples including the classification of skilled vs bad driver performance \cite{Hajinoroozi2016} and the prediction of performance in working memory tasks \cite{Johannesen2016}. However, while scalp EEG has proven effective at discerning the relevant brain activity changes that arise from changes in cognitive workload, it is obtrusive and thus impractical for daily life applications, while discrete and non-stigmatising wearable solutions are still being developed, such as Hearables \cite{Goverdovsky2017} and ear-EEG \cite{Looney2012} \cite{Nakamura2019}.

In recent years, eye gaze tracking has become a useful tool for tracking cognitive workload, with the ability to classify cognitive workload in real time in driving simulators \cite{Wang2018}, as well as predicting correctness in an N-back task whilst in a driving simulator \cite{VitoAmadori2020}. The ways to measure gaze and pupil dilation inevitably involve cameras; these are generally fixed and positioned to track the face and eyes. Such a gaze tracking apparatus is not wearable either, and is thus only applicable in scenarios such as driving or flying, where the camera can be mounted within the cockpit.

Other sensing modalities relevant for the estimation of cognitive load include electrocardiography (ECG) and photoplethsmography (PPG) and the use of the corresponding heart rate metrics to classify cognitive workload in a range of scenarios, including driving whilst performing an N-back memory task \cite{Tjolleng2017}, taking maths tests of varying difficulty \cite{Wang2019} and when engaging in a partially automated task with a machine based component \cite{Zhang2017}. Whilst ECG and PPG are less obtrusive than EEG or gaze tracking in daily life and do offer a wearable solution to cognitive workload tracking, it remains unclear whether the documented increases in heart rate are associated with the stress of performing well during higher cognitive workload tasks \cite{VonRosenberg2017} \cite{Chanwimalueang2017}, or indeed the increased cognitive workload itself. Namely, heart rate is known to correlate strongly with stress level whilst driving, as well as skin conductivity (sweat level) \cite{Healey2005}. For the purpose of exclusively measuring cognitive workload, it is therefore important to consider tasks whereby the aspect of stress that a maths test or driving may cause is negligible, whilst still maintaining the ability to vary cognitive workload.

\subsection{The brain, oxygen and cognitive workload}

The brain is the most metabolically active organ in the human body. At rest, the brain consumes 20\% of the body's oxygen \cite{Butterworth1999} and this percentage increases with increased cognitive demand. Oxygen restriction has significant effects on cognitive function; for example, less oxygen delivery to the brain has been observed in those with memory impairments \cite{Owen2011} \cite{Eustache1995}. Moreover, oxygen administration, through the breathing of supplemental oxygen and the associated increase in blood oxygen, has been shown to result in significantly better memory performance and reaction times \cite{Scholey1999} \cite{Moss1998} \cite{Kim2012} \cite{Chung2006}.

Functional near infrared spectroscopy (fNIRS), a tool for measuring oxygenation of tissue and thus oxygen consumption, has shown increases in oxygen consumption of the brain with an increase in cognitive workload in drivers \cite{Tsunashima2009}. Furthermore, fNIRS has helped to detect increased oxygenation of specific brain regions (such as the left inferior frontal gyrus, involved in language processing) with an increase in the difficulty of a letter based N-back memory task in pilots \cite{Ayaz2010}. This motivates us to investigate whether these changes in oxygen consumption are also observable in spectral analysis of blood, or manifest themselves through increases in breathing rate or breathing magnitude.

\subsection{In-ear SpO\textsubscript{2}}
The notion of blood oxygen saturation refers to the proportion of haemoglobin binding sites (four for each molecule) that are occupied with oxygen, out of the total number of haemoglobin binding sites available in the blood, and is formulated as:

\begin{equation}
    \text{Oxygen Saturation} = \frac{HbO_{2}}{HbO_{2} + Hb}
\end{equation}
where the symbol $Hb$ refers to haemoglobin not bound with oxygen and $HbO_{2}$ to haemoglobin bound to oxygen.

Arterial blood oxygen saturation is typically estimated using pulse oximetry to yield a percentage value (SpO\textsubscript{2}), with subjects that have a healthy respiratory system typically exhibiting SpO\textsubscript{2} values of 96-98\% at sea level \cite{Smith2012}. Photoplethysmography (PPG) is the non-invasive measurement of light absorption through the blood. PPG effectively measures the pulse, as with more blood present, less light is reflected, which yields a pulsatile increase in blood volume with each heart beat. Pulse oximetry uses PPG simultaneously at different wavelengths of light (red and infrared) to estimate blood oxygen levels. The level of blood oxygen saturation is mirrored in a change in the ratio of light absorbance between the red and infrared light. The extinction coefficient of oxygenated haemoglobin with red light ($\approx 660nm$) is lower than that of deoxygenated haemoglobin, and the reverse holds for infrared light ($ \approx 880-940nm$) \cite{Nitzan2014}. The simultaneous measurement of the absorbance/reflectance of both infrared and red light therefore allows for an estimation of blood oxygen saturation. 

Pulse oximetry can be measured at most skin sites, but for convenience it is usually measured from the finger. On the other hand, enthusiasm for wearable eHealth technology and Hearables \cite{Goverdovsky2017} has promoted the ear canal as a preferred site for the measurement of vital signs, given its proximity to the brain (ear-EEG \cite{Looney2012}) and the comparatively stable position of the head with respect to vital signs during daily life, in stark contrast with currently used sites such as the wrist.

In-ear pulse oximetry offers significant benefits over conventional finger pulse oximetry. Firstly, the in-ear location is robust to changes in blood volume which occur during vasoconstriction and hypothermia, giving a stable and accurate photoplethysmogram during cold exposure \cite{Budidha2015}. Next, it has been documented that in-ear PPG exhibits larger amplitude variations due to inspiration and expiration, allowing for a improved measurement of respiration rate \cite{Budidha2018}. Also, a significant delay has been evidenced between ear pulse oximetry and pulse oximetry on the right index finger \cite{Davies2020} and on the hand/foot \cite{Hamber1999} for detection of hypoxemia (low levels of blood oxygen). Indeed, SpO\textsubscript{2} from the right index finger was shown to take an average of 12.4 seconds longer to respond when compared with SpO\textsubscript{2} from the right ear canal during breath holds across different subjects \cite{Davies2020}. This is because the ear is supplied by the common carotid artery, in close proximity to the heart, making ear-SpO\textsubscript{2} an effective non-invasive proxy for priority core blood oxygen.

Given that the brain is the most metabolically active organ in the body and that it increases oxygen consumption with cognitive workload, we hypothesised that increased cognitive workload may be measurable through blood oxygen saturation. Considering that wearable in-ear pulse oximetry provides a robust SpO\textsubscript{2} signal with minimal delay, we set out to answer whether in-ear pulse oximetry can be used to accurately classify different levels of cognitive workload, and furthermore can this classification be performed in an almost real-time fashion?

\section{Methods}

\subsection{Hardware}

\begin{figure}[h]
\centerline{\includegraphics[width=0.5\textwidth]{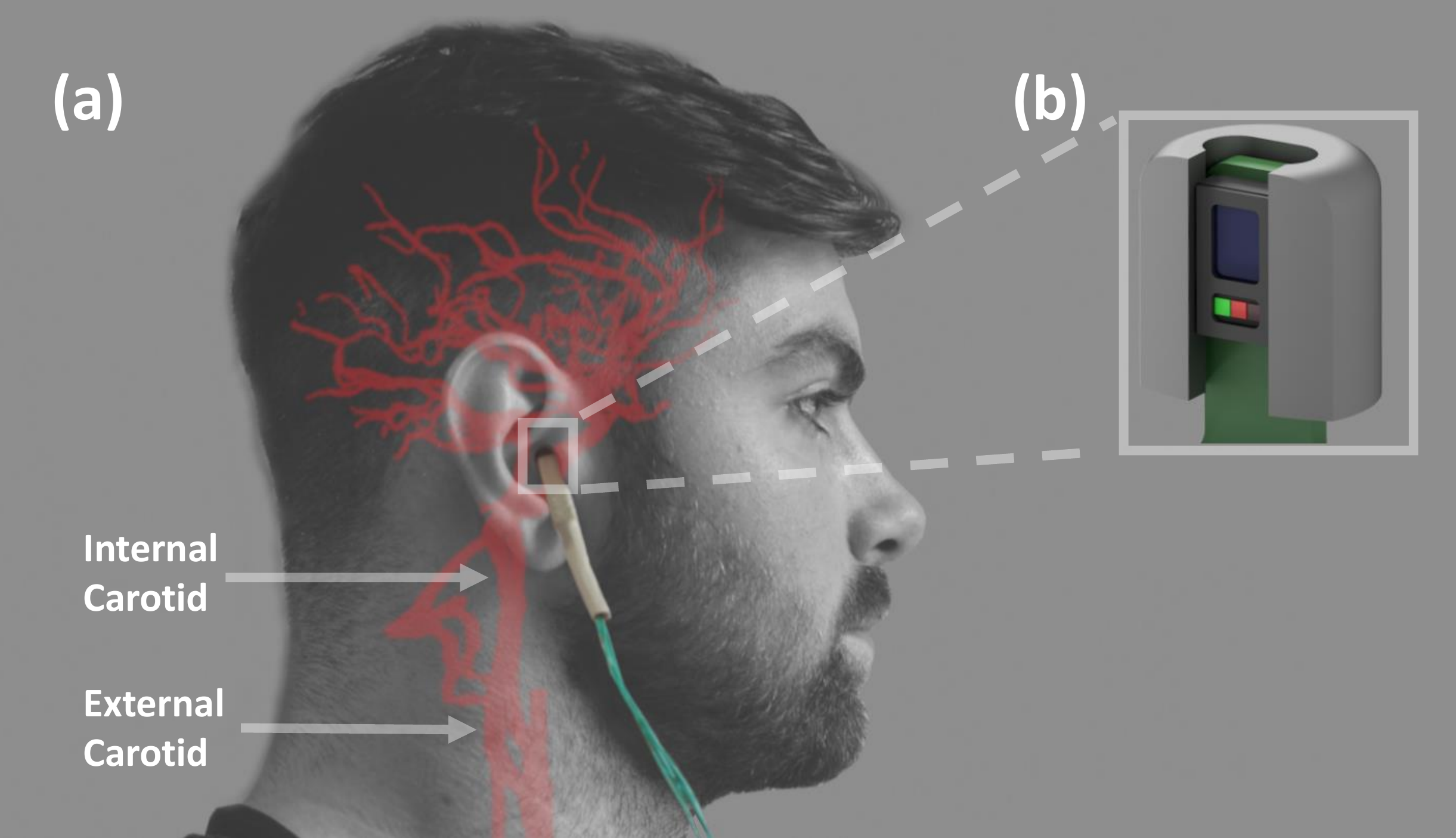}}
\caption{The in-ear photoplethysmography sensor used in our study. (a) The sensor placement within the ear canal, with the major arteries supplying the brain and the ear highlighted. (b) Zoom-in of the pulse oximetry sensor, with a form form factor of a viscoelastic memory foam earbud.}
\label{eareeg_3Dmodel}
\end{figure}

The photoplethysmography sensor used was the MAX30101 digital PPG chip by Maxim Integrated, consisting of green (537nm), red (660nm) and infrared (880nm) light emitting diodes as well as a photo-diode to measure the reflected light. In this study, the PPG sensor was embedded in a cut-out rectangular section of a viscoelastic foam earbud \cite{Goverdovsky2016}, allowing for comfortable insertion. The three light wavelengths (at 537nm, 660nm, 880nm) give three measures of pulse, and also the signals required to estimate blood oxygen saturation and respiration rate. The connected PPG circuitry, including decoupling capacitors and level shifting circuitry that enable digital communication between the 1.8 V and 3 V domains, was neatly covered with heat shrink and positioned just outside of the ear, as shown in Fig.~\ref{eareeg_3Dmodel}. This same sensor was secured to the right little finger with medical tape for finger PPG measurements when they were required. The sensor apparatus was connected to a circuit board which stores the data for each trial on an SD card.

MATLAB 2018a by MathWorks (Natick, MA, USA) was used to create a graphical user interface which refreshed four single digit numbers every 5 seconds on a screen in front of the subject. The MATLAB program also communicated with an Arduino Uno by Arduino (Somerville, MA, USA) with each refresh, which in turn communicated with the data logging circuit board with an electrical pulse to align the PPG data to each 5 second window.

\subsection{Experimental Protocol}
\begin{figure}[]
\centerline{\includegraphics[width=0.5\textwidth]{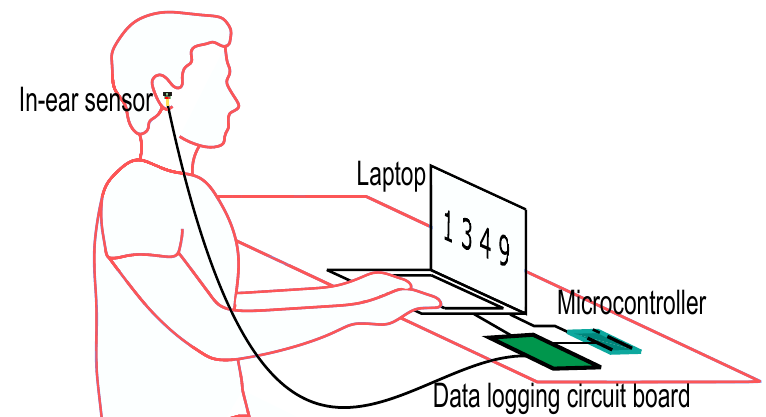}}
\caption{Illustration of the recording of in-ear SpO\textsubscript{2} during an N-back task. The SpO\textsubscript{2} sensor links to a circuit board which logs the data stream, as well as accepting input from the microcontroller. The 4 single digit numbers displayed on the laptop refresh every 5 seconds, and communicate this refresh time to the microcontroller, which in turn sends an electrical pulse to the circuit board to align the task with the physiological data.}
\label{recording_diagram}
\end{figure}
The participants in the recordings were 8 healthy subjects (4 males, 4 female) aged 21 - 29 years. A single PPG sensor was used per subject and was safely secured within the right ear canal; on 3 subjects recordings were repeated at a later date from the right little finger. The subjects were seated in front of a monitor during the recording where a MATLAB graphical user interface updated with 4 randomly generated single digit numbers every 5 seconds, as shown in Fig.~\ref{recording_diagram}. Subjects were asked to count the number of odd numbers and, depending on the N-back trial, they were tasked with entering the current number of odd numbers using the keyboard (0-back), the previous number (1-back), the number 2 steps back (2-back), or the number of odd numbers 3 steps back (3-back). Each trial lasted for 5 minutes and 40 seconds (68 5 second epochs), with 6 epochs used for calibration, leaving 62 epochs for analysis. Four trials performed by each subject, corresponding to the four levels of N-back task. Each subject was given between 5 and 10 minutes rest between trials, to ensure that trials could be considered independent of each other.

The recordings were performed under the IC ethics committee approval JRCO 20IC6414, and all subjects gave full informed consent.

\subsection{Signal Processing}

\begin{figure}[]
\centerline{\includegraphics[width=0.5\textwidth]{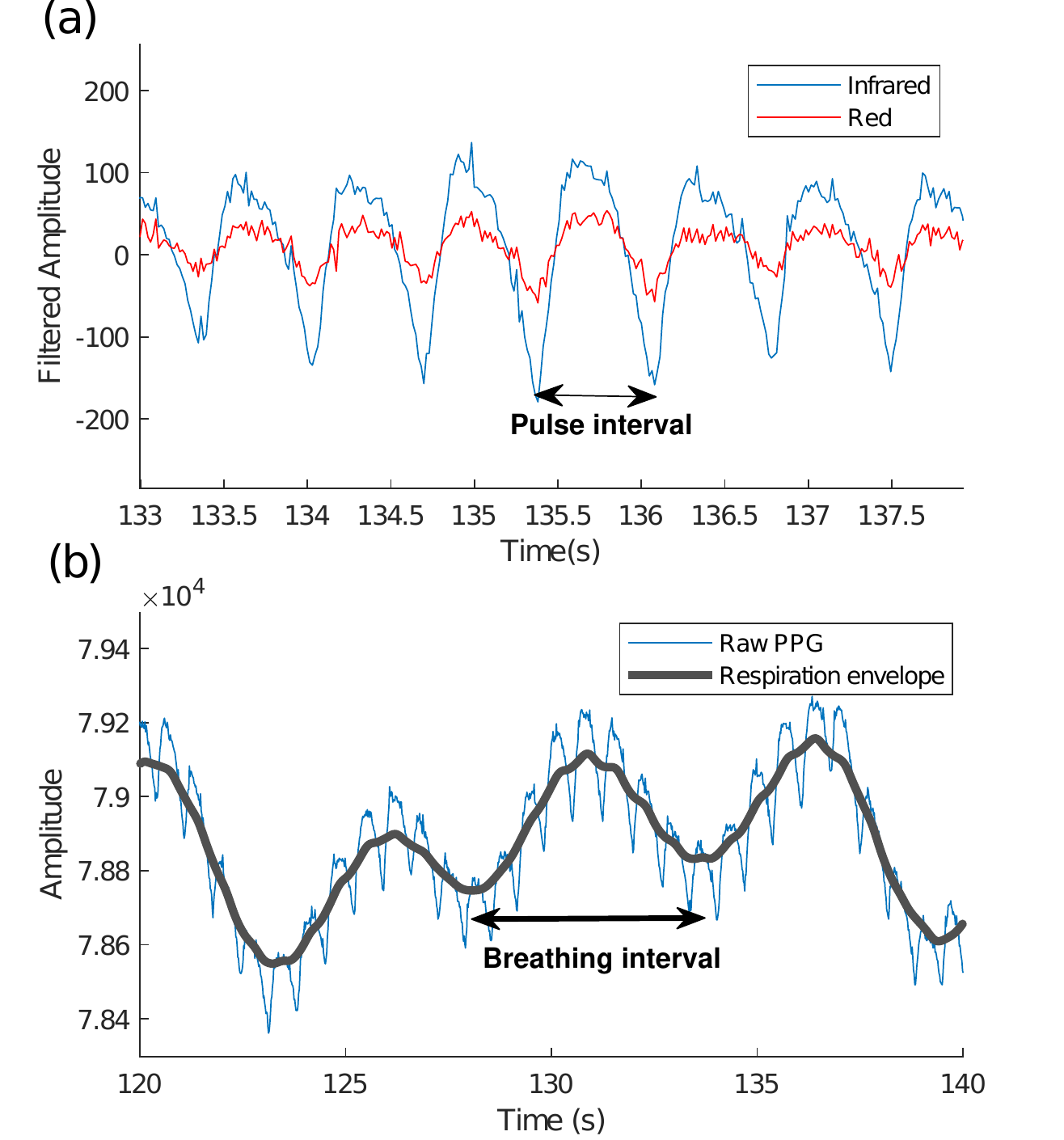}}
\caption{Calculation of the pulse interval and breathing interval based on in-ear PPG data. (a) Pulse interval calculation from a pre-filtered PPG signal. (b) Calculation of breathing interval based on the raw PPG signal, with large low frequency amplitude fluctuations visible (in grey) due to respiration.}
\label{pulse_and_breathing_cal}
\end{figure}

\begin{figure*}[]
\centerline{\includegraphics[width=1\textwidth]{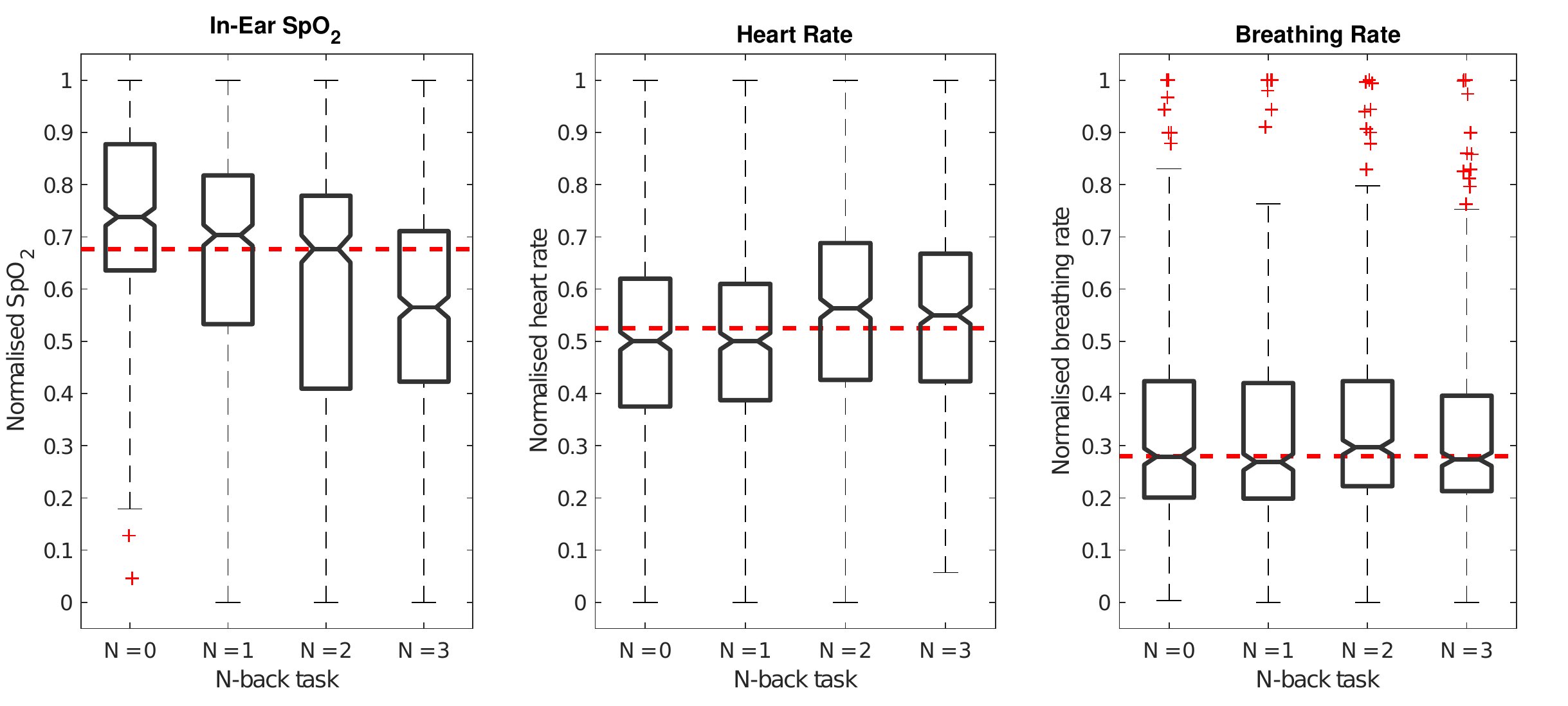}}
\caption{Box plots of SpO\textsubscript{2}, heart rate and breathing rate from the in-ear sensor for each 5 second epoch, normalised per subject and split into N-back categories. The top and bottom of each box represent respectively the upper and lower quartiles, the center notches of each box designate the median, and the whisker lines extending out of the box the range. Red crosses outside the range denote outliers. The broken red line across each plot represents the median across all N-back tasks.}
\label{normalised_collection}
\end{figure*}

\begin{figure}[]
\centerline{\includegraphics[width=0.31\textwidth]{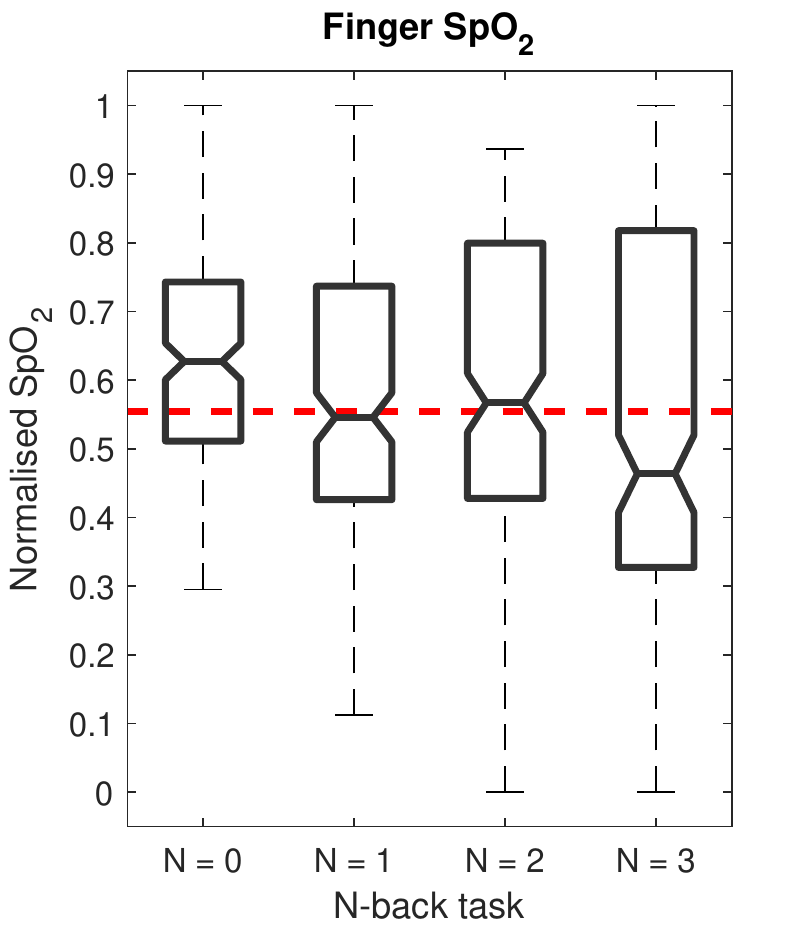}}
\caption{Box plots of SpO\textsubscript{2} from the finger sensor for each 5 second epoch, normalised per subject and split into N-back categories. The top and bottom of each box represent respectively the upper and lower quartiles, the center notches of each box designate the median, and the whisker lines extending out of the box the range. The broken red line across each plot represents the median across all N-back tasks.}
\label{finger_spo2}
\end{figure}

\begin{table}[] 
\caption{Summary of features used for classification} 
\centering 
\begin{tabular}{c| l} 
\hline\hline 
Category & Features\\ [0.25ex] 
\hline   
SpO\textsubscript{2} & SpO\textsubscript{2} mean, calibrated SpO\textsubscript{2} mean,\\ & red amplitude mean, infrared (IR) amplitude mean, \\
& calibrated red amplitude mean,\\
& calibrated IR amplitude mean, \\
& red AC/DC ratio mean, IR AC/DC ratio mean, \\
& red peak prominence mean, IR peak prominence mean,\\
& red/IR AC ratio mean,  red amplitude variance, \\
& IR amplitude variance. \\\hdashline
Pulse & Heart rate mean, calibrated heart rate mean, \\ 
& pulse full-width-half-maximum (FWHM) mean, \\
& pulse width ratio$^{\dagger}$ mean, pulse width ratio variance. \\ \hdashline
Breathing & Breathing rate mean, calibrated breathing rate mean\\
& breathing amplitude mean.\\\hline


\hline\hline  
\end{tabular} 
\label{summarytable} 

$^{\dagger}$ Pulse width ratio is the ratio between the FWHM of the peak and the FWHM of the trough, giving a systolic to diastolic duration ratio.
\end{table} 

\begin{figure}[]
\centerline{\includegraphics[width=0.5\textwidth]{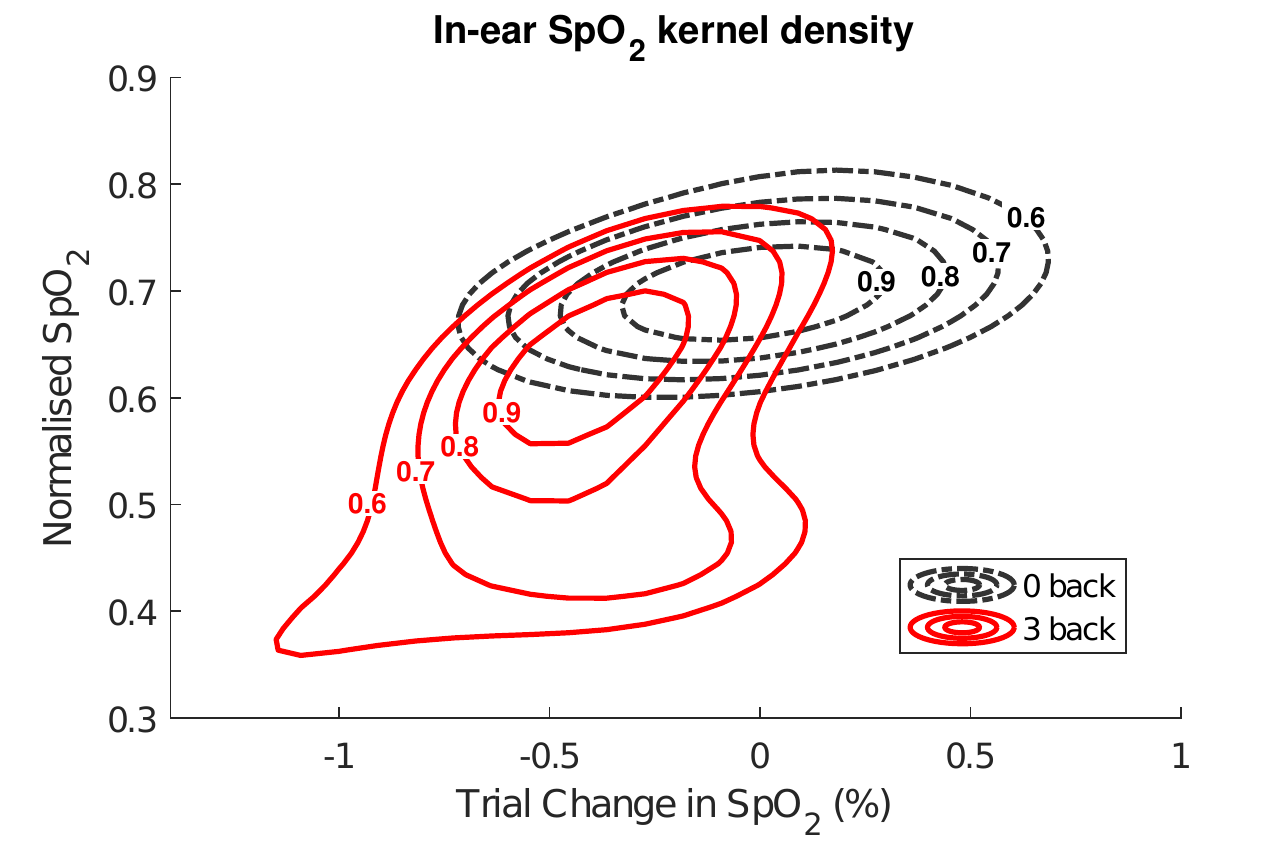}}
\caption{Two-dimensional contour plots of the kernel density estimates for mean SpO\textsubscript{2} normalised per subject, and trial change in SpO\textsubscript{2} (calibrated SpO\textsubscript{2}), derived from all of the 5 second epochs of data. The kernel density estimates are plotted independently for two categories: 0-back (black dotted line) and 3-back (red solid line). Kernel density was normalised between 0 and 1 for each category, and the corresponding values for each contour line shown are marked on the contour lines themselves.}
\label{kernel_density}
\end{figure}

\begin{figure*}[b]
\centerline{\includegraphics[width=1\textwidth]{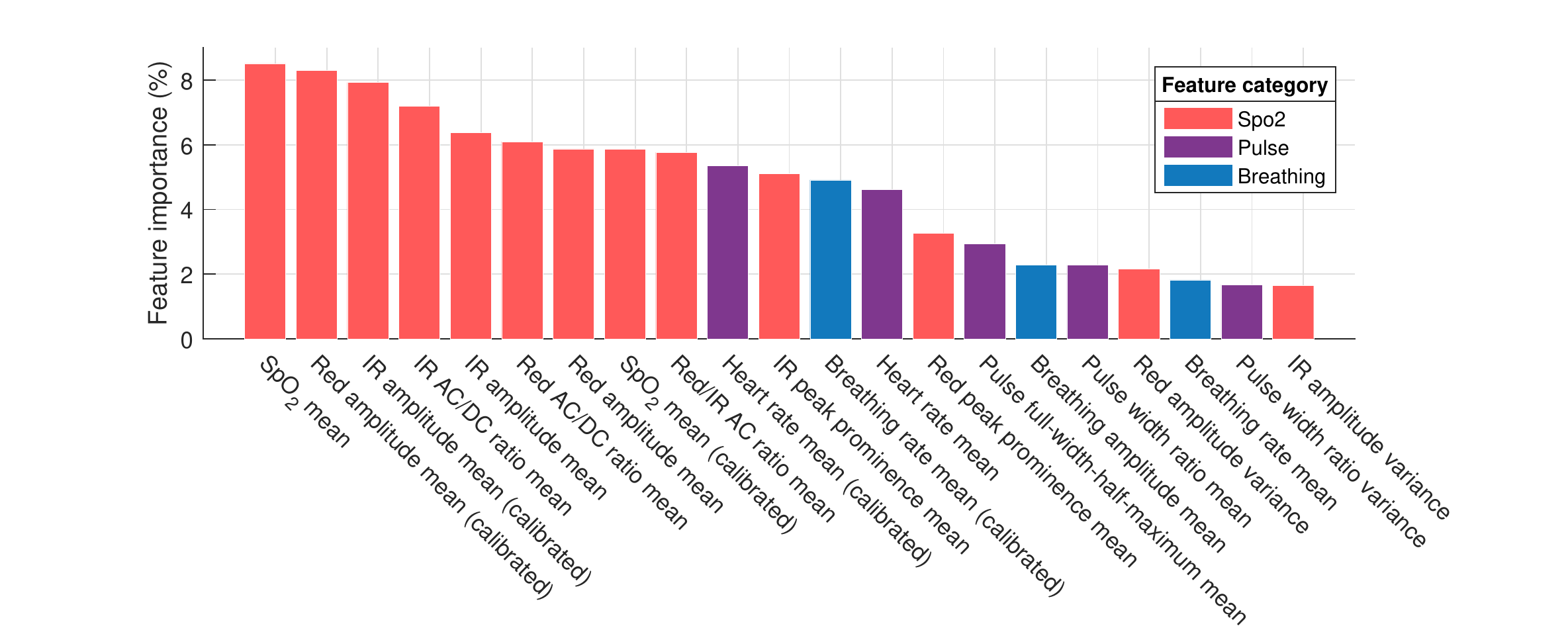}}
\caption{Feature importance for all 21 features considered. Feature importance was derived from the reduction in tree impurity based on the contribution of each feature to the random forest classifier. The features were split into 3 categories, depending on the physiological metric from which they were derived. The SpO\textsubscript{2} features are shown in red, pulse features are in purple, and breathing features are in blue.}
\label{feature_importance}
\end{figure*}

The ratio of absorbance of infra-red to red light within the PPG sensor changes depending on the proportion of haemoglobin that is oxygenated in the blood. This change can be quantified through the so called \textit{ratio of ratios} metric \cite{Rusch1996}, given by

\begin{equation}
    R = \frac{\frac{AC_{red}}{DC_{red}}}{\frac{AC_{infrared}}{DC_{infrared}}}
\end{equation}

An empirically derived linear approximation can then be used to calculate an SpO\textsubscript{2} value as a proxy to oxygen saturation. Using the manufacturers suggested calibration \cite{Maximintegrated2018}, the SpO\textsubscript{2} value was calculated as

\begin{equation}
    SpO_{2} = 104 - 17R
\end{equation}

To obtain the alternating current (AC) components within the PPG measurements, the raw signals were firstly band-pass filtered between 1Hz and 30Hz. Peak detection was performed using the MATLAB function \textit{findpeaks}, with a minimum peak prominence varied between 40 and 150 arbitrary units for the infrared signal and between 10 and 30 arbitrary units for the red signal. The same procedure was repeated for the inverted signals to find the troughs. Next, the peak values and trough values were separated and interpolated, before their absolute values were added together to give a constant estimate of the AC amplitude. The direct current (DC) components were obtained by low-pass filtering the raw signals at 0.01Hz.

The peak detection procedure of the AC infrared troughs was also used to calculate pulse rate, given that the PPG peak from the ear canal is broader than the peak from the finger (a characteristic of the pressure wave found in the carotid artery) and would thus give a noisy pulse rate estimate. An example of the pulse interval calculation is shown in Fig.~\ref{pulse_and_breathing_cal}a. 

Fluctuations in the baseline of ear-PPG due to inspiration and expiration have been evidenced as far stronger from the ear-canal than from the finger \cite{Budidha2018}. For the calculation of respiration rate, the raw PPG signal was first band-pass filtered between 0.2Hz and 30Hz, followed by a moving average filter with a 150 sample window (corresponding to 2.4 seconds). Peak detection was performed using the MATLAB function \textit{findpeaks} with a minimum peak prominence of 10, to give respiration peaks. The difference of the timings of these peaks was then used to give a breathing interval, shown in Fig.~\ref{pulse_and_breathing_cal}b. The inverse of the interval signal was then multiplied by 60 to give breathing rate (in breaths per minute). The amplitude values of the respiration peaks were also used as an estimate of breathing amplitude. No epochs of data were discarded, even in the presence of motion artifacts.

\subsection{Feature extraction}

For each 5 second epoch, 21 time domain features (13 SpO\textsubscript{2} based features, 5 pulse based features and 3 breathing based features) were extracted. Frequency-based features were not used as the 5 second window is too short for reliable heart rate variability metrics from PPG. The 21 features used summarised in Table~\ref{summarytable}.

\subsection{Classification and evaluation}

A random forest classifier with AdaBoost was employed from the publicly available scikit-learn Python toolbox \cite{Pedregosa2011}. For the random forest base, the number of trees was set to 100, the class weight was set to 'balanced subsample' and the maximum number of features was set to 5. For the AdaBoost framework, the random forest was set as the base classifier and the maximum number of estimators was set to 50.

Ten-fold cross-validation was employed on the fully shuffled data for the case of four-category classification (0-back, 1-back, 2-back, 3-back). Leave-one-subject-out cross-validation was employed on two-category classification (0-back and 3-back). Class-specific accuracy and overall accuracy were used as metrics to evaluate classification performance.

\section{Results}

The mean mistake percentages across subjects for each N-back stage were $2.1\%\pm 1.5\%$, $1.6\%\pm 1.6\%$, $13.1\%\pm 11.2\%$ and $29.7\%\pm 19.4\%$ for 0-back, 1-back, 2-back and 3-back tasks, respectively. The substantial increase in mistakes between 1-back, 2-back and 3-back tasks indicates that 3-back and 2-back were difficult enough to create a meaningful increase in cognitive workload.

\subsection{Change in blood oxygen, heart rate, breathing rate}

The mean recorded SpO\textsubscript{2} across all subjects and trials was $97.4 \pm 0.4\%$, the mean heart rate was $84.3 \pm 6.0$ beats per minute and the mean estimated breathing rate was $12.8 \pm 3.3$ breaths per minute. All results therefore fell into the physiologically expected range.

Overall, there was a decrease in the median normalised SpO\textsubscript{2} in the 3-back task compared with the easier 1-back and 2-back tasks. The 0-back task also had the highest median normalised SpO\textsubscript{2} out of all of the N-back tasks. These observations were present in both the ear and finger SpO\textsubscript{2} readings, as shown in Fig~\ref{normalised_collection} and Fig~\ref{finger_spo2}. Across all subjects, a slight increase in median normalised heart rate can be seen between low cognitive workload states (0-back and 1-back) and high cognitive workload states (2-back and 3-back). No significant change in estimated breathing rate is observed between trials. A one-way ANOVA was performed between N-back tasks the normalised ear-SpO\textsubscript{2}, normalised heart rate, and the normalised breathing rate. There was a significant difference between groups for ear-SpO\textsubscript{2} (F[3,1980] = 48.3, p = $3.89\times 10^{-30}$) and heart rate (F[3,1980] = 12.1, p = $7.36\times 10^{-8}$), but not breathing rate (F[3,1980] = 1.18, p = $0.31$).

In the in-ear recordings, the 3-back task led to either the lowest median absolute SpO\textsubscript{2} or largest median decrease in SpO\textsubscript{2} in 6 out of the 8 subjects, and in the remaining 2 subjects the 2-back task led to either the lowest median absolute SpO\textsubscript{2} or largest median decrease in SpO\textsubscript{2}. Highest median heart rate was observed in the 3-back task in 3 out of 8 subjects, and highest median breathing rate was observed in the 3-back task in 2 out of the 8 subjects. The substantial changes in median SpO\textsubscript{2} normalised per subject across N-back tasks can be seen in Fig~\ref{normalised_collection}.

Fig~\ref{kernel_density} demonstrates the high separability of 0-back and 3-back tasks, through a two-dimensional kernel density plot of two features, namely SpO\textsubscript{2} normalised on a per subject basis, and trial change in SpO\textsubscript{2}.

\subsection{Classification}

\subsubsection{Shuffled ten-fold cross-validation}

\begin{figure}[]
\centerline{\includegraphics[width=0.38\textwidth]{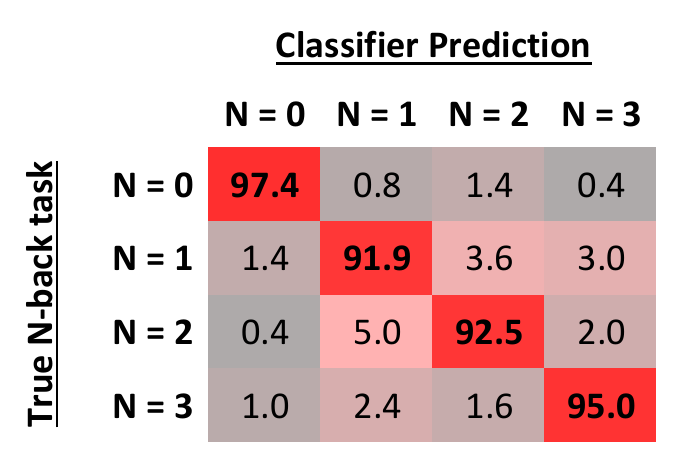}}
\caption{Mean confusion matrix for the results of ten-fold shuffled cross-validation in four-category prediction (0-back, 1-back, 2-back and 3-back). The rows correspond to the true N-back category, and the columns to the category predicted by the classifier. }
\label{conf_matrix}
\end{figure}

With ten-fold cross-validation, shuffled across all participants, we were able to classify the 5-second 0-back epochs with an average accuracy of 97.4\%, 1-back epochs with an accuracy of 91.9\%, 2-back with 92.5\% and 3-back with 95.0\%, giving a total average classification accuracy of 94.2\%. The largest errors occurred with the miss-classification of 1-back as 2-back and vice versa, which was expected given their medium difficulty natures. Classification accuracy was notably better on 0-back and 3-back tasks. Fig~\ref{conf_matrix} shows the full confusion matrix averaged across 10-fold cross validation. Averaged feature importance (according to reduction in tree impurity in the random forest) across each fold indicates that out of 21 features, the 9 most important features for classification were all SpO\textsubscript{2} derived features and mean SpO\textsubscript{2} was the most important feature. The complete list of employed features together with their corresponding feature importance, is shown in Fig~\ref{feature_importance}. The 10-fold cross validation was also implemented on the smaller 3-subject finger data set, which achieved an mean overall accuracy of 85.9\%.

\subsubsection{Leave-one-subject-out cross-validation}

\begin{figure}[h]
\centerline{\includegraphics[width=0.5\textwidth]{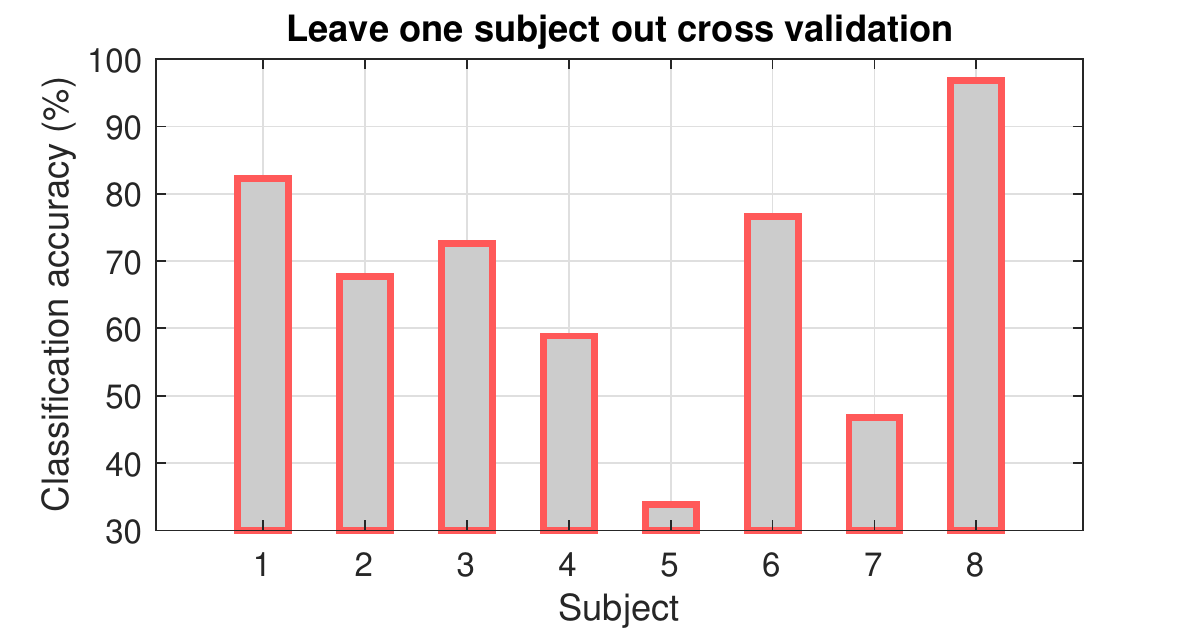}}
\caption{Classification accuracy for testing on each subject, with the classifier trained exclusively on the other 7 subjects.}
\label{leaveonesubject}
\end{figure}

The utilisation of leave-one-subject-out training on just the 0-back and 3-back data had very variable results, ranging from 34\% accuracy at worst, to 97\% accuracy at best. The accuracy percentages for testing on each subject are shown in Fig~\ref{leaveonesubject}. The mean overall accuracy was 66.9\%. 

\section{Discussion}
As hypothesised, increased cognitive workload in general led to a decrease in SpO\textsubscript{2}, and this decrease is likely due to increased oxygen consumption by the brain. This change occurs in both the ear and finger sensors, and in our experiments it was more notable from the ear. This indicates that the in-ear SpO\textsubscript{2} sensor indeed offers more sensitivity to the small fluctuations in blood oxygen that occur due to changes in cognitive workload. Furthermore, the mean SpO\textsubscript{2} was either the lowest or had the largest decrease across the trial in tasks with high cognitive workload, in all 8 subjects. This demonstrates the robustness of the SpO\textsubscript{2} response to changes in cognitive workload, compared with the commonly used metric of heart rate where increased heart rate did not necessarily correspond to increased cognitive workload.

The 5-second epochs of in-ear data achieved an average classification accuracy of 94.2\% across the four N-back task categories. Out of the 21 time domain features used, the best 9 features in terms of contribution to classification accuracy were all SpO\textsubscript{2} based features. In our data, blood oxygen levels were thus a far better predictor of cognitive workload than pulse or breathing rate. This is further indicated by the large separability of kernel density estimates, plotted in Fig~\ref{kernel_density}. These results indicate that in-ear SpO\textsubscript{2} represents a valuable new modality for making possible real-time cognitive workload classification with high accuracy.

There was a large variability in performance in leave-one-subject-out training and testing using the two categories of 0-back and 3-back, ranging from 34\% to 97\% accuracy. In general, cognitive workload tasks induce different levels in cognitive workload in different people, which can be seen in the large standard deviation in mistakes. The physiological response to increased cognitive workload also varies widely between different people. Given these factors, it was not expected that leave-one-subject-out training would work well given the relatively small sample size of 8 subjects. The ability of leave-one-subject-out training to perform well when testing on a few of the 8 subjects shows firstly that there is a strong similarity in the SpO\textsubscript{2} response to changes in cognitive workload that becomes visible even across a few subjects. Secondly, it indicates that there is promise with a larger training pool for a classifier to effectively generalise across subjects using in-ear SpO\textsubscript{2}, and thus accurately classify cognitive workload of unseen subjects in real time.

\section{Conclusion}

We have introduced and validated a proof of concept for cognitive workload estimation using a novel wearable in-ear SpO\textsubscript{2} sensor. Pulse oximetry from the ear canal has been shown to be capable of discriminating between 4 categories of cognitive workload based on an N-back task over 5 second epochs, with a mean accuracy of 94.2\%. High cognitive workload in the 2-back and 3-back tasks has led to either the lowest median absolute SpO\textsubscript{2} or largest median decrease in SpO\textsubscript{2} in all of the subjects, therefore demonstrating a robust decrease in blood oxygen in response to increased cognitive workload. This decrease in SpO\textsubscript{2} was plausibly caused by the increased oxygen consumption of the brain under increased cognitive demands. Similar trends in SpO\textsubscript{2} were also observed from the standard finger pulse oximetry recordings, however finger SpO\textsubscript{2} was found not to be as effective at classifying into the 4 categories, likely due to the previously documented large and varying delay in detection of blood oxygen at the finger. Overall, this study has established in-ear SpO\textsubscript{2} as an effective tool for classification of cognitive workload, to be used alone or in combination with current gold standard workload tracking equipment such as EEG and ECG, or within the emerging multi-modal Hearables.

\section*{Acknowledgment}
This work was supported by the Racing Foundation grant 285/2018, MURI/EPSRC grant EP/P008461 and the Dementia Research Institute at Imperial College London.

\FloatBarrier



\begin{thebibliography}{10}
\providecommand{\url}[1]{#1}
\csname url@samestyle\endcsname
\providecommand{\newblock}{\relax}
\providecommand{\bibinfo}[2]{#2}
\providecommand{\BIBentrySTDinterwordspacing}{\spaceskip=0pt\relax}
\providecommand{\BIBentryALTinterwordstretchfactor}{4}
\providecommand{\BIBentryALTinterwordspacing}{\spaceskip=\fontdimen2\font plus
\BIBentryALTinterwordstretchfactor\fontdimen3\font minus
  \fontdimen4\font\relax}
\providecommand{\BIBforeignlanguage}[2]{{%
\expandafter\ifx\csname l@#1\endcsname\relax
\typeout{** WARNING: IEEEtran.bst: No hyphenation pattern has been}%
\typeout{** loaded for the language `#1'. Using the pattern for}%
\typeout{** the default language instead.}%
\else
\language=\csname l@#1\endcsname
\fi
#2}}
\providecommand{\BIBdecl}{\relax}
\BIBdecl

\bibitem{Ayaz2010}
H.~Ayaz, B.~Willems, B.~Bunce, P.~Shewokis, K.~Izzetoglu, S.~Hah, A.~Deshmukh,
  and B.~Onaral, ``{Cognitive workload assessment of air traffic controllers
  using optical brain imaging sensors},'' \emph{Advances in Understanding Human
  Performance: Neuroergonomics, Human Factors Design, and Special Populations},
  pp. 21--31, 2010.

\bibitem{Hajinoroozi2016}
M.~Hajinoroozi, Z.~Mao, T.~P. Jung, C.~T. Lin, and Y.~Huang, ``{EEG-based
  prediction of driver's cognitive performance by deep convolutional neural
  network},'' \emph{Signal Processing: Image Communication}, vol.~47, pp.
  549--555, Sep 2016.

\bibitem{Johannesen2016}
J.~K. Johannesen, J.~Bi, R.~Jiang, J.~G. Kenney, and C.-M.~A. Chen, ``{Machine
  learning identification of EEG features predicting working memory performance
  in schizophrenia and healthy adults},'' \emph{Neuropsychiatric
  Electrophysiology}, vol.~2, no.~1, p.~3, Dec 2016.

\bibitem{Goverdovsky2017}
V.~Goverdovsky, W.~{Von Rosenberg}, T.~Nakamura, D.~Looney, D.~J. Sharp,
  C.~Papavassiliou, M.~J. Morrell, and D.~P. Mandic, ``{Hearables: Multimodal
  physiological in-ear sensing},'' \emph{Scientific Reports}, vol.~7, no.~1,
  pp. 1--10, Dec 2017.

\bibitem{Looney2012}
D.~Looney, P.~Kidmose, C.~Park, M.~Ungstrup, M.~Rank, K.~Rosenkranz, and
  D.~Mandic, ``{The in-the-ear recording concept: User-centered and wearable
  brain monitoring},'' \emph{IEEE Pulse}, vol.~3, no.~6, pp. 32--42, 2012.

\bibitem{Nakamura2019}
T.~Nakamura, Y.~D. Alqurashi, M.~J. Morrell, and D.~Mandic, ``{Hearables:
  Automatic overnight sleep monitoring with standardised in-ear EEG sensor},''
  \emph{IEEE Transactions on Biomedical Engineering}, pp. 1--1, 2019.

\bibitem{Wang2018}
R.~Wang, P.~V. Amadori, and Y.~Demiris, ``{Real-time workload classification
  during driving using HyperNetworks},'' in \emph{IEEE International Conference
  on Intelligent Robots and Systems}.\hskip 1em plus 0.5em minus 0.4em\relax
  Institute of Electrical and Electronics Engineers Inc., Dec 2018, pp.
  3060--3065.

\bibitem{VitoAmadori2020}
P.~{Vito Amadori}, T.~Fischer, R.~Wang, and Y.~Demiris, ``{Decision
  anticipation for driving assistance systems},'' in \emph{IEEE International
  Conference on Intelligent Transportation Systems (ITSC 2020)}, vol. 2020,
  2020.

\bibitem{Tjolleng2017}
A.~Tjolleng, K.~Jung, W.~Hong, W.~Lee, B.~Lee, H.~You, J.~Son, and S.~Park,
  ``{Classification of a driver's cognitive workload levels using artificial
  neural network on ECG signals},'' \emph{Applied Ergonomics}, vol.~59, pp.
  326--332, Mar 2017.

\bibitem{Wang2019}
C.~Wang and J.~Guo, ``{A data-driven framework for learners' cognitive load
  detection using ECG-PPG physiological feature fusion and XGBoost
  classification},'' in \emph{Procedia Computer Science}, vol. 147.\hskip 1em
  plus 0.5em minus 0.4em\relax Elsevier B.V., Jan 2019, pp. 338--348.

\bibitem{Zhang2017}
J.~Zhang, S.~Li, and R.~Wang, ``{Pattern recognition of momentary mental
  workload based on multi-channel electrophysiological data and ensemble
  convolutional neural networks},'' \emph{Frontiers in Neuroscience}, vol.~11,
  no. MAY, p. 310, May 2017.

\bibitem{VonRosenberg2017}
W.~von Rosenberg, T.~Chanwimalueang, T.~Adjei, U.~Jaffer, V.~Goverdovsky, and
  D.~P. Mandic, ``{Resolving Ambiguities in the LF/HF Ratio: LF-HF Scatter
  Plots for the Categorization of Mental and Physical Stress from HRV},''
  \emph{Frontiers in Physiology}, vol.~8, no. JUN, p. 360, Jun 2017.

\bibitem{Chanwimalueang2017}
T.~Chanwimalueang, L.~Aufegger, T.~Adjei, D.~Wasley, C.~Cruder, D.~P. Mandic,
  and A.~Williamon, ``{Stage call: Cardiovascular reactivity to audition stress
  in musicians},'' \emph{PLOS ONE}, vol.~12, no.~4, p. e0176023, Apr 2017.

\bibitem{Healey2005}
J.~A. Healey and R.~W. Picard, ``{Detecting stress during real-world driving
  tasks using physiological sensors},'' \emph{IEEE Transactions on Intelligent
  Transportation Systems}, vol.~6, no.~2, pp. 156--166, Jun 2005.

\bibitem{Butterworth1999}
R.~F. Butterworth, \emph{{Hypoxic Encephalopathy}}, 6th~ed., A.~R. {Siegel GJ},
  Agranoff~BW, Ed.\hskip 1em plus 0.5em minus 0.4em\relax Lippincott-Raven,
  1999.

\bibitem{Owen2011}
L.~Owen and S.~I. Sunram-Lea, ``{Metabolic agents that enhance ATP can improve
  cognitive functioning: A review of the evidence for glucose, oxygen,
  pyruvate, creatine, and l-carnitine},'' \emph{Nutrients}, vol.~3, no.~8, pp.
  735--755, Aug 2011.

\bibitem{Eustache1995}
F.~Eustache, P.~Rioux, B.~Desgranges, G.~Marchal, M.~C. Petit-Tabou{\'{e}},
  M.~Dary, B.~Lechevalier, and J.~C. Baron, ``{Healthy aging, memory subsystems
  and regional cerebral oxygen consumption},'' \emph{Neuropsychologia},
  vol.~33, no.~7, pp. 867--887, Jul 1995.

\bibitem{Scholey1999}
A.~B. Scholey, M.~C. Moss, N.~Neave, and K.~Wesnes, ``{Cognitive performance,
  hyperoxia, and heart rate following oxygen administration in healthy young
  adults},'' \emph{Physiology and Behavior}, vol.~67, no.~5, pp. 783--789, Nov
  1999.

\bibitem{Moss1998}
M.~C. Moss, A.~B. Scholey, and K.~Wesnes, ``{Oxygen administration selectively
  enhances cognitive performance in healthy young adults: A placebo-controlled
  double-blind crossover study},'' \emph{Psychopharmacology}, vol. 138, no.~1,
  pp. 27--33, 1998.

\bibitem{Kim2012}
S.~P. Kim, M.~H. Choi, J.~H. Kim, H.~W. Yeon, H.~J. Yoon, H.~S. Kim, J.~Y.
  Park, J.~H. Yi, G.~R. Tack, and S.~C. Chung, ``{Changes of 2-back task
  performance and physiological signals in ADHD children due to transient
  increase in oxygen level},'' \emph{Neuroscience Letters}, vol. 511, no.~2,
  pp. 70--73, Mar 2012.

\bibitem{Chung2006}
S.~C. Chung, J.~H. Sohn, B.~Lee, G.~R. Tack, J.~H. Yi, J.~H. You, J.~H. Jun,
  and R.~Sparacio, ``{The effect of transient increase in oxygen level on brain
  activation and verbal performance},'' \emph{International Journal of
  Psychophysiology}, vol.~62, no.~1, pp. 103--108, Oct 2006.

\bibitem{Tsunashima2009}
H.~Tsunashima and K.~Yanagisawa, ``{Measurement of brain function of car driver
  using functional near-infrared spectroscopy (fNIRS)},'' \emph{Computational
  Intelligence and Neuroscience}, vol. 2009, 2009.

\bibitem{Smith2012}
G.~B. Smith, D.~R. Prytherch, D.~Watson, V.~Forde, A.~Windsor, P.~E. Schmidt,
  P.~I. Featherstone, B.~Higgins, and P.~Meredith, ``{SpO2 values in acute
  medical admissions breathing air-Implications for the British Thoracic
  Society guideline for emergency oxygen use in adult patients?}''
  \emph{Resuscitation}, vol.~83, no.~10, pp. 1201--1205, Oct 2012.

\bibitem{Nitzan2014}
M.~Nitzan, A.~Romem, and R.~Koppel, ``{Pulse oximetry: Fundamentals and
  technology update},'' pp. 231--239, Jul 2014.

\bibitem{Budidha2015}
K.~Budidha and P.~A. Kyriacou, ``{Investigation of photoplethysmography and
  arterial blood oxygen saturation from the ear-canal and the finger under
  conditions of artificially induced hypothermia},'' in \emph{Proceedings of
  the Annual International Conference of the IEEE Engineering in Medicine and
  Biology Society, EMBS}, vol. 2015-Novem.\hskip 1em plus 0.5em minus
  0.4em\relax Institute of Electrical and Electronics Engineers Inc., Nov 2015,
  pp. 7954--7957.

\bibitem{Budidha2018}
------, ``{In vivo investigation of ear canal pulse oximetry during
  hypothermia},'' \emph{Journal of Clinical Monitoring and Computing}, vol.~32,
  no.~1, pp. 97--107, Feb 2018.

\bibitem{Davies2020}
H.~J. Davies, I.~Williams, N.~S. Peters, and D.~P. Mandic, ``{In-ear SpO2: A
  tool for wearable, unobtrusive monitoring of core blood oxygen saturation},''
  \emph{Sensors}, vol.~20, no.~17, p. 4879, Aug 2020.

\bibitem{Hamber1999}
E.~A. Hamber, P.~L. Bailey, S.~W. James, D.~T. Wells, J.~K. Lu, and N.~L. Pace,
  ``{Delays in the detection of hypoxemia due to site of pulse oximetry probe
  placement},'' \emph{Journal of Clinical Anesthesia}, vol.~11, no.~2, pp.
  113--118, Mar 1999.

\bibitem{Goverdovsky2016}
V.~Goverdovsky, D.~Looney, P.~Kidmose, and D.~P. Mandic, ``{In-ear EEG from
  viscoelastic generic earpieces: Robust and unobtrusive 24/7 monitoring},''
  \emph{IEEE Sensors Journal}, vol.~16, no.~1, pp. 271--277, Jan 2016.

\bibitem{Rusch1996}
T.~L. Rusch, R.~Sankar, and J.~E. Scharf, ``{Signal processing methods for
  pulse oximetry},'' \emph{Computers in Biology and Medicine}, vol.~26, no.~2,
  pp. 143--159, 1996.

\bibitem{Maximintegrated2018}
MaximIntegrated, ``{Recommended configurations and operating profiles for
  MAX30101/MAX30102 EV Kits},'' Tech. Rep., 2018.

\bibitem{Pedregosa2011}
F.~Pedregosa, G.~Varoquaux, A.~Gramfort, V.~Michel, B.~Thirion, O.~Grisel,
  M.~Blondel, P.~Prettenhofer, R.~Weiss, V.~Dubourg, J.~Vanderplas, A.~Passos,
  D.~Cournapeau, M.~Brucher, M.~Perrot, and {\'{E}}.~Duchesnay,
  ``{Scikit-learn: Machine Learning in Python},'' \emph{Journal of Machine
  Learning Research}, vol.~12, no. Oct, pp. 2825--2830, 2011.

\end{thebibliography}
\end{document}